\documentclass{revtex4}
\usepackage{graphics,amsmath}

\newcount\n  \newdimen\w

\def\Repeat#1#2{\n=#1\relax\loop\ifnum       
  \n>0\relax #2\advance\n by-1\repeat}

\long\def\OMIT#1{\relax }  

\def\re#1{(\ref{#1})}   
  
\def\eqn#1#2{ \begin{align} \label{#1}         #2 \end{align}}

\def\nl#1{          \\ \label{#1}        }  
\def\nnl#1{ \tag*{} \\ \label{#1}        }  
\def\nln#1{         \\ \label{#1} \tag*{}}  

\def\delim#1#2#3{\csname\ifcase#1 relax\or   
   big\or Big\or bigg\or Bigg\fi\endcsname   
  {\ifcase#2\or\Delim#3\or\deliM#3\fi}}      
\def\Delim#1{\ifcase#1\relax\or(\or[\or\{\or<\or\langle\or|\or\|\or---{ }\fi}
\def\deliM#1{\ifcase#1\relax\or)\or]\or\}\or>\or\rangle\or|\or\|\or{ }---\fi}

\def\largerfrac#1#2#3{      
  \whichtypesize\n=\currenttypesize\advance\n by #1 \mathchoice
  {\setbox0\hbox{$\displaystyle-$} \w=.5\ht0\advance\w by-.5\dp0\setbox0
    \hbox{\typesize\n $\displaystyle-$} \advance\w by -.5\ht0\advance\w
    by .5\dp0\raise\w \hbox{\typesize\n$\displaystyle{\frac{#2}{#3}}$}}
  {\setbox0\hbox{$-$} \w=.5\ht0 \advance\w by -.5\dp0 \setbox0\hbox
    {\typesize\n $-$} \advance\w by-.5\ht0\advance\w by
    .5\dp0\raise\w\hbox{\typesize\n$\frac{#2}{#3}$}}
  {\setbox0\hbox{$\scriptstyle-$} \w=.5\ht0 \advance\w by-.5\dp0\setbox0
    \hbox{\typesize\n $\scriptstyle-$} \advance\w by -.5\ht0 \advance\w
    by .5\dp0 \raise\w\hbox{\typesize\n$\scriptstyle{\frac{#2}{#3}}$}}
  {\setbox0\hbox{$\scriptscriptstyle-$} \w=.5\ht0
    \advance\w by -.5\dp0 \setbox0\hbox{\typesize\n
    $\scriptscriptstyle-$} \advance\w by -.5\ht0 \advance\w by .5\dp0
    \raise\w\hbox{\typesize\n$\scriptscriptstyle{\frac{#2}{#3}}$}}  }

\def\d{{\rm d}}       

\begin{document}

\title{Non-equilibrium thermodynamics and Newtonian gravitation}
\author{P. V\'an$^{1,2,3}$}
\affiliation{$^1$Department of Theoretical Physics, Wigner Research Centre for Physics, H-1525 Budapest, Konkoly Thege Mikl\'os u. 29-33., Hungary; \\
$^2$Department of Energy Engineering, Faculty of Mechanical Engineering,  Budapest University of Technology and Economics, 1111 Budapest, Műegyetem rkp. 3., Hungary;\\
$^3$Montavid Thermodynamics Research Group, Hungary}
\author{S. Abe}
\affiliation{Department of Physical Engineering, Mie University, Mie 514-8507, Japan}
 
\date{\today}

\begin{abstract}
Classical gravitation is treated from the point of view of non-equilibrium thermodynamics. Gravitational potential is a thermodynamic state variable in a weakly nonlocal treatment. Entropy production is calculated and the simplest solution of the inequality with corresponding fluxes and forces is given. In case of ideal gravitation without dissipation one obtains the Poisson equation.
\end{abstract}


\maketitle


Thermodynamics and gravitation are both universal theories, because their validity is independent of  material structure and composition. Thermodynamics can be connected to gravitation when both are considered as emergent theories. Emergent in the sense that their laws are originated in an underlying microscopic dynamics. The various attempts to connect thermodynamics and gravitation are focusing on general relativity (some of the many notable attempts are \cite{Jac95a,EliEta06a,Ver11a,Ver17a}). According to this concept, Newtonian gravitation must be emergent in the same sense. In this work we embed Newtonian gravitation into nonequilibrium thermodynamics. With our approach the emergence of gravitation is based on the entropy inequality alone, without any particular microscopic background. However, a particular aspect of the thermodynamic theory is weak nonlocality, it is based on gradient dependent thermodynamic potentials, where the field energy of gravitation plays an important role. 

In nonrelativistic continua the gravitational force density is a source term in the momentum balance, the gravitational energy is represented by the potential and is separated from the internal energy. The power of gravitation can be smaller or larger than zero, because it is considered as an external effect. Gravity is introduced through the gravitational potential, which is subtracted from the total energy density (see e.g. \cite{GroMaz62b,Gya70b}). The distinguishing property of our approach is the separate representation of the field energy of gravitation \cite{Syn72a,FraSza11a,DewWea18a}. This field energy is proportional to the square gradient of the potential, therefore the equation of state depends on the gradient of a state variable, therefore it is a weakly nonlocal contribution from the thermodynamic point of view.

Weakly nonlocal extensions frequently appear in physics. Notable examples are the phase field theories \cite{Waa894a,LanKha54a}, but generalized continua, see e.g. \cite{CosCos09b,Eri99b},  are also weakly nonlocal from a thermodynamic point of view, and can be treated in a non-equilibrium thermodynamic framework, \cite{VanEta14a}. A correct weakly nonlocal extension of non-equilibrium thermodynamics requires a concept of extensivity, with an extension of the classical first order Euler homogeneity to gradient dependent thermodynamic potentials \cite{BerVan17b}. In this paper we introduce a theory of gravitation in the framework of weakly nonlocal non-equilibrium thermodynamics and obtain the Poisson equation in the ideal, nondissipative case, when direct gravitation related dissipation is zero. First, thermostatics is treated, then the entropy production is calculated. For the sake of simplicity only heat conducting fluids are considered, the extension to solids and coupling to other classical fields is straightforward. 

\section{Thermostatics of gravitation}

In the following thermostatic relations are expressed with specific quantities. The specific volume $v$ is related to the mass density, as $\rho = 1/v$. The specific total energy is denoted by $e$, and the specific energy of gravitating matter, the gravitational potential  by $\varphi$. Then the internal energy $u$ is the difference of the usual internal energy $e$ and the energies of the gravitating matter and field. With specific quantities it is given in the following form
\eqn{generg}{
u=e - \varphi - \frac{\nabla\varphi\cdot\nabla\varphi}{8\pi G\rho}. 
} 
Here $\nabla$ is the spatial derivative, $\cdot$ denotes the contraction. $G$ is the gravitational constant. The last term is the specific gravitational field energy, whose separation from the internal energy is the key aspect of the theory. Then the specific entropy is the function of $u$ and the specific volume $v=1/\rho$, and the Gibbs relation becomes
\eqn{gravGrel}{
\d u = T\d s - p\d v = \d e - \d \varphi - \d\left( \frac{\nabla\varphi\cdot\nabla \varphi}{8\pi G\rho}\right).
}

Here $T$ denotes  the temperature and $p$ is the thermostatic pressure. The specific quantities are convenient for the following calculations. The extensivity of the matter and field, that is the first order Euler homogeneity of the entropy is expressed with the previous specific quantities as 
\eqn{gravext}{
u = Ts - pv + \mu = e - \varphi -  \frac{\nabla\varphi\cdot\nabla\varphi}{8\pi G\rho}.
}
Here $\mu$ is the chemical potential and this formula is the consequence of the extensivity of the gravitating matter and field.

\section{Entropy production}

The entropy balance expresses the second law in the form of the following inequality
\eqn{bal_entr}{
\rho\dot s + \nabla\cdot \mathbf{J} = \sigma \geq 0.
}

Here $J$ is the entropy flux, comoving current density of the of the entropy. The specific entropy is a function of the specific internal energy $u$, the density $\rho$, the gravitational potential $\varphi$ and its gradient $\nabla\varphi$.  The entropy inequality is conditional, the balances of mass and internal energy are to be applied as constraints. 

The first constraint is the conservation of mass, which is given by the equation
\eqn{bal_mass}{
\dot \rho + \rho\nabla\cdot {\bf v} = 0.
}

Here the dot denotes the substantial or comoving time derivative, that is $\dot a = \frac{\partial a}{\partial t} + {\bf v}\cdot \nabla a $ for any $a(t,{\bf x})$ nonrelativistic fields. Furthermore ${\bf v}$ is the velocity field of the continuum, defined in the usual way as mass and momentum flow \cite{VanEta17a}. The balance of momentum is given as
\eqn{bal_mom}{
\rho\dot {\bf v} + \nabla \cdot \mathbf{P} = \mathbf{0}.
}
Here $\mathbf{P}$ is the pressure tensor, which is symmetric as we assume that the internal rotation is moment of momentum of the continuum is zero. The balance of internal energy follows as (see e.g. \cite{Gya70b})
\eqn{bal_inte}{
\rho\dot e + \nabla \cdot \mathbf{q} = - \mathbf{P}:\nabla\mathbf{v},
}
where $\mathbf{q}$ is the heat flux, the conductive current density of the internal energy, and the double dot denotes the trace of the product of the corresponding tensors $\mathbf{P}:\nabla\mathbf{v} = Tr(\mathbf{P}\cdot\nabla\mathbf{v})$ .

With the help of the Gibbs relation and the balances of mass and internal energy, \re{bal_mass} and \re{bal_inte} it is straightforward to calculate the entropy balance. In the Appendix we have given a simple derivation, based on separation of divergences, generalizing de Groot and Mazur for weakly nonlocal state spaces \cite{GroMaz62b,Mau06a,Van18bc}. Then the entropy balance is obtained in the following form:
\eqn{entrbaltot}{
\rho\dot s &+\nabla\cdot\left[\frac{1}{T}\left(\mathbf{q}+\frac{1}{4\pi G}          \dot\varphi\nabla\varphi\right)\right] = \nln{k1} 
	&\left(\mathbf{q}+\frac{\dot{\varphi}}{4\pi G} \nabla\varphi\right)\cdot\nabla\left(\frac{1}{T}\right) + 	\frac{\dot\varphi}{4\pi G T}\left(\Delta\varphi - 4\pi G\rho\right) - \nln{k2}
	&\left[\mathbf{P} - p\mathbf{I} - \frac{1}{4\pi G}\left(\nabla\varphi\nabla\varphi -
		\frac{1}{2}\nabla\varphi\cdot\nabla\varphi \mathbf{I} \right) \right]:\frac{\nabla \mathbf{v}}{T} \geq 0,
}
where $\mathbf{I}$ is the second order unit tensor, and the entropy flux is ${\bf J} =\frac{1}{T}\left(\mathbf{q}+\frac{1}{4\pi G} \dot\varphi\nabla\varphi\right)$. 


Now one  obtains the usual solution of the entropy inequality after the identification of thermodynamic forces and fluxes. In our case the heat flux $\mathbf{q}$ and the pressure tensor $\mathbf{P}$ are constitutive quantities and the classical thermodynamic forces of thermal and mechanical interactions can be identified \cite{GroMaz62b}. Also, according to contemporary theories of nonequilibrium thermodynamics the evolution equation of the gravitational potential $\dot \varphi$ can be considered as a constitutive function \cite{VanAta08a,BerVan17b}. The corresponding  thermodynamic fluxes and forces are given in Table \ref{ff}.
\begin{table}
\centering
\begin{tabular}{c|c|c|c}
       &Thermal  & Gravitational &  Mechanical \\ \hline
Fluxes & $\mathbf{q}+\frac{\dot{\varphi}}{4\pi G} \nabla\varphi$ & 
		 $\dot\varphi$ & 
	 	 $\mathbf{P} - p\mathbf{I} - \mathbf{P}_{grav}$ \\ \hline
Forces & $\nabla \left(\frac{1}{T}\right)$ & 
		 $\frac{1}{T}\left(\frac{\Delta\varphi }{4\pi G}- \rho\right)$ & 
	 	 $-\frac{\nabla \mathbf{v}}{T}$\\ 
\end{tabular}\\
\caption{Thermodynamic fluxes and forces of self-gravitating fluids. $\mathbf{P}_{grav}$ is the gravitational pressure.}
\label{ff}
\end{table}

Both the mechanical and thermal thermodynamic fluxes have changed due to the presence of gravitation. There is a contribution to the heat flux and also to the pressure, $\mathbf{P}_{grav}= \frac{1}{4\pi G}\left(\nabla\varphi\nabla\varphi -\frac{1}{2}\nabla\varphi\cdot\nabla\varphi\mathbf{I} \right)$. The difference of the thermal flux from the heat flux is usual in case of weakly nonlocal internal variables \cite{Van03a}. The thermal interaction is vectorial, the mechanical is second order tensorial and the gravitational is scalar. Therefore, according to the representation theorems of isotropic materials, which is the Curie principle in our case, cross effects are possible only between the scalar part of the pressure and the gravitation. Hence the linear constitutive equations for the heat flux and the pressure are
\eqn{constrel}{
&\mathbf{q} + \frac{\dot{\varphi}}{4\pi G} \nabla\varphi = 
\lambda \nabla \left(\frac{1}{T}\right) =  -\lambda_F \nabla T,  \nl{cr_heat}
&Tr(\mathbf{P}) - 3 p +\frac{\nabla\varphi\cdot\nabla\varphi}{8\pi G } = \nnl{e2} 
	&\qquad\frac{l_{21}}{T} \left(\frac{\Delta\varphi}{4\pi G } -\rho\right) - \frac{l_2}{T} \nabla\cdot\mathbf{v}, \nl{cr_bpress}
&\mathbf{P}\! -\! Tr(\mathbf{P})\frac{\mathbf{I}}{3}\! -\! \frac{1}{4\pi G}\left(\nabla\varphi\nabla\varphi \!-\! \frac{1}{3}(\nabla\varphi)^2\mathbf{I}\right) =	\nnl{[e3]}
	&\qquad-\eta\left(\nabla \mathbf{v}\! + (\nabla{\bf v})^T\! -\! \frac{2}{3}\nabla\cdot\mathbf{v}\mathbf{I}\right)\!. 
}

Here the upper index $T$ denotes the transpose of the related tensor, $\lambda_F = \lambda T^{-2}$ is the Fourier heat conduction coefficient, $\frac{l_2}{3T} = \eta_v$ is the bulk viscosity and $\eta$ is the shear viscosity. 

Moreover, the differential equation for the gravitational potential is 
\eqn{cr_grav}{
\dot\varphi = \frac{l_1}{T} \left(\frac{\Delta\varphi}{4\pi G } -\rho\right) - \frac{l_{12}}{T} \nabla\cdot\mathbf{v}.
}

This is our main result. 

The transport coefficients $\lambda, \eta_v, \eta, l_1, l_2$ are non negative and $L = l_1 l_2 - (l_{12}+l_{21})^2/4\geq 0$, according to the entropy inequality. 

\subsection{Ideal gravitation} 
If the Poisson equation, $\Delta\varphi - 4\pi G\rho = 0$,  is satisfied then the thermodynamic force of- gravitation is zero. In this case the divergence of the gravitational pressure is proportional to the gravitational force:
\eqn{gpress}{
\nabla \cdot \mathbf{P}_{grav} = \nabla\cdot\left[ \frac{1}{4\pi G}\left(\nabla\varphi\nabla\varphi -\frac{1}{2}\nabla\varphi\cdot\nabla\varphi\mathbf{I} \right)\right] = \rho \nabla\varphi.
}
Hence we obtain the usual balances  of self-gravitating fluids, where there is only thermal and mechanical dissipation and the balance of momentum can be written as usual:
$$
\rho \dot {\bf v} + \nabla\cdot \mathbf{P}  = -\rho \nabla\varphi.
$$

The dissipation is zero, if the heat flux and the viscous pressure are both zero, that is $\mathbf{q}=0$ and $\mathbf{\Pi}  = \mathbf{P} - p\mathbf{I} = 0$. 

However, the most general nondissipative self-gravitating fluid is more general. If the coupling between the gravitational and bulk mechanical interactions is antisymmetric, the entropy production is zero if $\lambda, \eta, \eta_v,l_1,l_2=0$. In this case denoting $l_{12} = -l_{21} = \hat l \neq 0$, the constitutive functions are given as
\eqn{ig1}{
\dot \varphi &= -\hat l \nabla\cdot\mathbf{v}, \nl{ig2}
Tr(\mathbf{P})    &= 3p- \frac{(\nabla\varphi)^2}{8\pi G } - \frac{\hat l}{T} \left(\frac{\Delta\varphi}{4\pi G } - \rho\right).
}

\section{Discussion}

We have shown that Newtonian gravitation emerges in the framework of non-equilibrium thermodynamics if the gravitational potential $\varphi$ is a thermodynamic state variable, as in recent thermostatic theories of nonextensive thermostatistics \cite{LatPer13a}. One obtains the Poisson equation if the contribution of gravitation to the entropy production is zero. If the field energy density is quadratic in the gradient of the potential -- as it is most natural -- then nonequilibrium thermodynamics leads to a dissipative theory of self-gravitating fluids where the thermodynamic force of the gravitational interaction is proportional to the Poisson equation.

The well known shape and form dependence of extended gravitating bodies is a particular form of nonextensivity, which is attributed to long range forces \cite{LatPer13a,Cha06a}. The weak nonlocality, the presence of the field energy of gravitation in the thermodynamic potentials is introduced and meaningful only in a local treatment, for the entropy density or for the specific entropy. For an extended thermodynamic body with spatially varying gravitational field it becomes naturally nonextensive, in the sense that the body potential is first order Euler homogeneous function of the volume of the body only for homogeneous gravitational fields. Thermodynamics of field theories, including self gravitating ones, is best constructed from a local, density based approach.

Any speculation about gravity as a thermodynamic related emergent phenomenon has to deal with the Newtonian theory. The presented approach is universal, the second law of thermodynamics and the conservation of mass, momentum and energy are the conditions. Therefore it is a framework of any possible microscopic interpretation. The consequent structural aspects turned out to be unusual, but somehow straightforward in a continuum treatment. For example the time dependence of the gravitational potential follows from the coupling with other fields also in the nondissipative case. Then it is clear, that the identification of additional time dependent phenomena and their separation from other continuum couplings is an open experimental problem \cite{YanEta15m,Dio13a}.

Gravitation, as far as we know, is not dissipative.
However, there are several conceptual and practical problems and applications that can be treated naturally in a dynamic and  thermodynamic framework with ideal gravitation, too. For example the role of field energy in gravothermal effects, in particular gravothermal instabilities \cite{LynWoo68a,Thi70a,Pad90a,Lyn99a}, can influence the stability of static equilibria. Recent continuum treatments, like \cite{SorBer13a,GioEta19m}, with a nonequilibrium thermodynamic extension may improve our understanding of this phenomena. 

It is also remarkable, that in our case the weak equivalence principle, the equivalence of inertial and gravitational mass is a consequence of using the same mass density in the field energy term of \re{generg} and in the balance of momentum \re{bal_mom}. In this way the equivalence principle follows if the field energy of gravitation, $\rho_{gravf} = \frac{\nabla\varphi\cdot\nabla\varphi}{8\pi G}$ is independent of the mass density, or vice versa the equivalence principle requires a mass independent field energy density. 
 
A further property of our approach is that the treatment of other kind of gravitating continua, like self gravitating elastic solids and coupling to other interactions is also simple and straightforward, including more complex rheoelastic or multicomponent media \cite{BerVan17b}. 
Finally we would like to stress, that the presented treatment is nonrelativistic but reference frame independent as one can see from a Galilean relativistic treatment of fluids in the nonrelativistic spacetime \cite{Van17a}. This is because a gradient of a scalar is a spacelike covector, invariant of the change of reference frames. 

\subsection{Acknowledgement}   
{\bf Acknowledgement} The work was supported by the grants National Research, Development and Innovation Office – NKFIH​116197(116375), NKFIH 124366(124508) and NKFIH 123815. The authors thank to Robert Trasarti-Battistoni and Tam\'as F\"ul\"op for valuable discussions.

\section{Appendix}

{\bf Appendix.} Here we show a simple calculation of the entropy balance \re{entrbaltot}. The key aspect of the derivation is the Gibbs relation \re{gravGrel}, which enhances the  separation of surface and bulk contributions, according to the classical method of de Groot and Mazur \cite{GroMaz62b}. Let us remark, that there are weakly nonlocal generalizations of the more rigorous Coleman-Noll or Liu methods for exploiting the entropy inequality \cite{Van05a,Cim07a}. 

In this calculation we calculate the comoving time derivative of the entropy density and then substitute the balances of mass, \re{bal_mass}, and internal energy, \re{bal_inte}: 
\begin{widetext}
\eqn{bigcalc}{
\rho\dot s &= 
	\rho \left(\partial_e s \dot e + \partial_\rho s \dot \rho + \partial_\varphi s\dot \varphi + \partial_{\nabla\varphi} (\nabla\varphi)^\cdot\right) = \nnl{l1} 
	&=\frac{1}{T}\rho \dot e + \frac{p}{T}\rho \dot v - \frac{\rho}{T}\dot{\varphi} - \frac{1}{8\pi G T\rho}(\nabla\varphi)^2\dot \rho - \frac{1}{4\pi G T}\nabla\varphi\cdot(\nabla\varphi)^\cdot =\nnl{l2}
	&=-\nabla\cdot\left(\frac{\mathbf{q}}{T}\right) + \mathbf{q}\cdot\nabla\left(\frac{1}{T}\right) -\frac{1}{T} \mathbf{P}:\nabla\mathbf{v} + 
		 \frac{p}{T}\nabla\cdot{\bf v} -\frac{\rho}{T}\dot{\varphi} - \frac{1}{8\pi G T}(\nabla\varphi)^2\nabla\cdot\mathbf{v} -\nnl{l3}
		&\qquad- \nabla\cdot\left(\frac{1}{4\pi G T}\dot\varphi\nabla\varphi\right) +  
		\frac{\dot{\varphi}}{4\pi G} \nabla\varphi\cdot\nabla\left(\frac{1}{T}\right)+
		\frac{\dot\varphi}{4\pi G T}\Delta\varphi + 
		\frac{1}{4\pi G T}\nabla\varphi\cdot\nabla \mathbf{v} \cdot\nabla\varphi= \nnl{k5} 
	&=-\nabla\cdot\left[\frac{1}{T}\left(\mathbf{q}+\frac{1}{4\pi G} \dot\varphi\nabla\varphi\right)\right] 
		+\left(\mathbf{q}+\frac{\dot{\varphi}}{4\pi G} \nabla\varphi\right)\cdot\nabla\left(\frac{1}{T}\right) + 	\frac{\dot\varphi}{4\pi G T}\left(\Delta\varphi - 4\pi G\rho\right) - \nnl{bc}
	 &	\qquad -\left[\mathbf{P} - p\mathbf{I} - \frac{1}{4\pi G}\left(\nabla\varphi\nabla\varphi -\frac{1}{2}\nabla\varphi\cdot\nabla\varphi \mathbf{I} \right) \right]:\frac{\nabla \mathbf{v}}{T} \geq 0. 
}
\end{widetext}

Here we have used that the gradient and the substantial time derivative do not commute, because
$$
(\nabla\varphi)^\cdot = \nabla \dot{\varphi} -\nabla\varphi\cdot\nabla\mathbf{v}.
$$
Now, one can identify the entropy flux in the last two lines of \re{bc} as full divergence and the entropy production as follows.


\begin{thebibliography}{10}

\bibitem{Jac95a}
T.~Jacobson.
\newblock Thermodynamics of spacetime: The {E}instein equation of state.
\newblock {\em Physical Review Letters}, 75:1260--3, 1995.

\bibitem{EliEta06a}
C.~Eling, R.~Guedens, and T.~Jacobson.
\newblock Nonequilibrium thermodynamics of spacetime.
\newblock {\em Physical Review Letters}, 96(12):121301, 2006.

\bibitem{Ver11a}
E.~P. Verlinde.
\newblock On the origin of gravity and the laws of {N}ewton.
\newblock {\em Journal of High Energy Physics}, 2011(04):029, 2011.

\bibitem{Ver17a}
E.~P. Verlinde.
\newblock {Emergent Gravity and the Dark Universe}.
\newblock {\em SciPost Phys.}, 2:016, 2017.

\bibitem{GroMaz62b}
S.~R. de~Groot and P.~Mazur.
\newblock {\em Non-equilibrium Thermodynamics}.
\newblock North-Holland Publishing Company, Amsterdam, 1962.

\bibitem{Gya70b}
I.~Gyarmati.
\newblock {\em Non-equilibrium Thermodynamics /{F}ield Theory and Variational
  Principles/}.
\newblock Springer Verlag, Berlin, 1970.

\bibitem{Syn72a}
J.L. Synge.
\newblock Newtonian gravitational field theory.
\newblock {\em Il Nuovo Cimento B}, 8(2):373--390, 1972.

\bibitem{FraSza11a}
J.~Frauendiener and L.~B. Szabados.
\newblock A note on the post-newtonian limit of quasi-local energy expressions.
\newblock {\em Classical and Quantum Gravity}, 28(23):235009, 2011.

\bibitem{DewWea18a}
N.~Dewar and J.~O. Weatherall.
\newblock On gravitational energy in Newtonian theories.
\newblock {\em Foundations of Physics}, 48(5):558--578, 2018.

\bibitem{Waa894a}
J.~D. Van~der Waals.
\newblock Thermodynamische theorie der {K}apillarit\"at unter {V}oraussetzung
  stetiger {D}ichteänderung.
\newblock {\em Zeitschrift f\"ur Physikalische Chemie}, 13:657--725, 1894.

\bibitem{LanKha54a}
L.~D. Landau and I.~M. Khalatnikov.
\newblock Ob anomal'nom pogloshchenii zvuka vblizi tochek fazovogo perekhoda
  vtorogo roda.
\newblock {\em Dokladu Akademii Nauk, SSSR}, 96:469--472, 1954.
\newblock English translation: On the anomalous absorption of sound near a
  second order transition point. in: Collected papers of L. D. Landau, ed. D.
  ter Haar,(Pergamon, Oxford, 1965), pp. 626-633.

\bibitem{CosCos09b}
E.~Cosserat and F.~Cosserat.
\newblock {\em Th\'eorie des Corps D\'eformables}.
\newblock Hermann and Fils, Paris, 1909.

\bibitem{Eri99b}
C.~Eringen.
\newblock {\em Microcontinuum Field Theories I. Foundations and Solids}.
\newblock Springer-Verlag, Berlin-etc.., 3th edition, 1999.

\bibitem{VanEta14a}
P.~V\'an, C.~Papenfuss, and A.~Berezovski.
\newblock Thermodynamic approach to generalized continua.
\newblock {\em Continuum Mechanics and Thermodynamics}, 25(3):403--420, 2014.
\newblock Erratum: 421-422.

\bibitem{BerVan17b}
A.~Berezovski and P.~V\'an.
\newblock {\em Internal Variables in Thermoelasticity}.
\newblock Springer, 2017.

\bibitem{VanEta17a}
P.~V\'an, M.~Pavelka, and M.~Grmela.
\newblock Extra mass flux in fluid mechanics.
\newblock {\em Journal of Non-Equilibrium Thermodynamics}, 42(2):133--151,
  2017.

\bibitem{Mau06a}
G.~A. Maugin.
\newblock On the thermomechanics of continuous media with diffusion and/or weak
  nonlocality.
\newblock {\em Archive of Applied Mechanics}, 75:723--738, 2006.

\bibitem{Van18bc}
P.~V\'an.
\newblock Weakly nonlocal non-equilibrium thermodynamics: the {Cahn-Hilliard}
  equation.
\newblock In {\em Generalized Models and Non-classical Approaches in Complex
  Materials 1}, pages 745--760. Springer, 2018.

\bibitem{VanAta08a}
P.~V\'an, A.~Berezovski, and J.~Engelbrecht.
\newblock Internal variables and dynamic degrees of freedom.
\newblock {\em Journal of Non-Equilibrium Thermodynamics}, 33(3):235--254,
  2008.

\bibitem{Van03a}
P.~V\'an.
\newblock Weakly nonlocal irreversible thermodynamics.
\newblock {\em Annalen der Physik (Leipzig)}, 12(3):146--173, 2003.

\bibitem{LatPer13a}
I. Latella and A. P{\'e}rez-Madrid.
\newblock Local thermodynamics and the generalized {G}ibbs-{D}uhem equation in systems with long-range interactions.
\newblock {\em Physical Review E}, 88(4):042135, 2013.

\bibitem{Cha06a}
P-H. Chavanis.
\newblock Phase transitions in self-gravitating systems.
\newblock {\em International Journal of Modern Physics B}, 20(22):3113--3198,  2006.

\bibitem{YanEta15m}
Huan Yang, L.~R. Price, N.~D. Smith, R.~X. Adhikari, Haixing Miao, and Yanbei
  Chen.
\newblock Towards the laboratory search for space-time dissipation.
\newblock {\em arXiv:1504.02545}, 2015.

\bibitem{Dio13a}
L.~Di{\'o}si.
\newblock Note on possible emergence time of {N}ewtonian gravity.
\newblock {\em Physics Letters A}, 377(31-33):1782--1783, 2013.

\bibitem{LynWoo68a}
D.~Lynden-Bell and R.~Wood.
\newblock The gravo-thermal catastrophe in isothermal spheres and the onset of
  red-giant structure for stellar systems.
\newblock {\em Monthly Notices of the Royal Astronomical Society}, 138(4):495--525, 1968.

\bibitem{Thi70a}
W.~Thirring.
\newblock Systems with negative specific heat.
\newblock {\em Zeitschrift f{\"u}r Physik A Hadrons and nuclei}, 235(4):339--352, 1970.

\bibitem{Pad90a}
T.~Padmanabhan.
\newblock Statistical mechanics of gravitating systems.
\newblock {\em Physics Reports}, 188(5):285--362, 1990.

\bibitem{Lyn99a}
D.~Lynden-Bell.
\newblock Negative specific heat in astronomy, physics and chemistry.
\newblock {\em Physica A}, 263(1-4):293--304, 1999.

\bibitem{SorBer13a}
M.C. Sormani and G.~Bertin.
\newblock Gravothermal catastrophe: The dynamical stability of a fluid model.
\newblock {\em Astronomy \& Astrophysics}, 552:A37, 2013.

\bibitem{GioEta19m}
D.~Giordano, P.~Amodio, F.~Iavernaro, A.~Labianca, M.~Lazzo, F.~Mazzia, and L.~Pisani.
\newblock Fluid statics of a self-gravitating perfect-gas isothermal sphere.
\newblock {\em arXiv:1903.04044}, 2019.

\bibitem{Van17a}
P.~V\'an.
\newblock Galilean relativistic fluid mechanics.
\newblock {\em Continuum Mechanics and Thermodynamics}, 29(2):585--610, 2017.

\bibitem{Van05a}
P.~V\'an.
\newblock Exploiting the {S}econd {L}aw in weakly nonlocal continuum physics.
\newblock {\em Periodica Polytechnica, Ser. Mechanical Engineering},
  49(1):79--94, 2005.

\bibitem{Cim07a}
V.~A. Cimmelli.
\newblock An extension of {L}iu procedure in weakly nonlocal thermodynamics.
\newblock {\em Journal of Mathematical Physics}, 48:113510, 2007.

\end{thebibliography}

\end{document}